\documentclass{article}

\usepackage[a4paper, total={6in, 8in}]{geometry}
\usepackage[utf8]{inputenc}
\usepackage[T1]{fontenc}
\usepackage{enumitem}
\usepackage{pifont} 
\usepackage{amsmath}
\usepackage{amssymb}
\usepackage{csquotes}
\usepackage{graphicx}
\usepackage{xcolor}
\usepackage{framed}
\usepackage{cite}
\usepackage{palatino}
\usepackage[pdftex,bookmarks,colorlinks,linkcolor=blue,citecolor=blue, urlcolor=blue]{hyperref}

\title{\bf Emergent dynamical phases and collective motion in termites}
\author{Leticia R. Paiva$^{1}$, Sidiney G. Alves$^{1}$, Og DeSouza$^{2}$, Octavio Miramontes$^{3,*}$\\ \\$^{1}$ Departamento de F\'isica e Matem\'atica, \\Universidade Federal de S\~ao Jo\~ao Del-Rei 36420-000, Ouro Branco, MG, Brazil\\
$^{2}$Laboratorio de Termitologia, Universidade Federal de Vi\c cosa, \\36570-900 Vi\c cosa, Minas Gerais, Brazil\\
$^{3}$Departamento de Sistemas Complejos, Instituto de Física,\\ Universidad Nacional
Autónoma de México,\\ Ciudad de México, C.P. 04510, Mexico\\ $^{*}$octavio@fisica.unam.mx}
\date{\today}

\begin{document}

\maketitle

\begin{abstract}
\noindent Termites which are able to forage in the open can be often seen, in the field or in the lab: (i) wandering around, forming no observable pattern, or (ii) clustering themselves in a dense and almost immobile pack, or (iii) milling about in a circular movement. Despite been well reported patterns, they are normally regarded as independent phenomena whose specific traits have never been properly quantified. Evidence, however, favours the hypothesis that these are interdependent patterns, arisen from self-organised interactions and movement among workers. After all, termites are a form of active matter where blind cooperative individuals are self-propelled and lack the possibility of visual cues to spatially
orientate and align. It follows that their non-trivial close-contact patterns could generate motion-collision induced phase separations. This would then trigger the emergence of these three patterns (disorder, clustering, milling) as parts of the same continuum. By inspecting termite groups confined in arenas, we could quantitatively describe each one of these patterns in detail. We identified disorder, clustering and milling spatial patterns. These phases and their transitions are characterised aiming to offer refinements in the understanding of these aspects of self-propelled particles in active matter where close-range contacts and collisions are important.
\end{abstract}

\section{Introduction}

Active matter are systems made of a large number of interacting constituents (alive or not) able to convert some source of energy into directed motion \cite{ramaswamy2010mechanics,de2015introduction,fodor2018}. In addition to being self-propelled, these constituents interact among themselves in such a way that their movement is both synchronized and correlated \cite{czirok1997spontaneously,czirok2000collective}. 
In doing so, collective behaviour and pattern formation spontaneously emerges. Pattern formation is a form of spatio-temporal self-organisation that is ubiquitous in nature, spanning physical, chemical, biological and even social phenomena \cite{cross1993pattern}. 
In living matter, it is regarded as a fundamental out-of-equilibrium process underlying morphogenesis at the cellular level \cite{gierer1972theory, wolpert1978pattern, koch1994biological, isaeva2012self}, to quote an example. However it is also present as a product of the complex collective interactions of individuals in social groups. Swarms, fish schools and insect foraging trails are examples {\it par excellence} \cite{attanasi2014finite, chowdhury2004self, camazine2020self}.  The distinct ordered and disordered phases exhibited by various artificial systems composed of active particles have been investigated \cite{Magistris, fodor2018}. For example, patterns found by evolving the original Vicsek model \cite{Vicsek} present either a disordered phase or an ordered flocking state. For $N\rightarrow \infty$, in particular, a coexistence of two different phases is observed close to criticality \cite{Chate}.

\begin{figure}[!ht]
\begin{center}
\includegraphics*[width=0.95\textwidth]{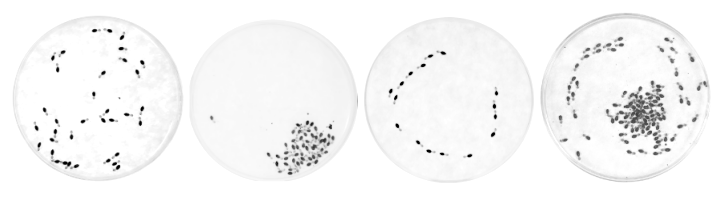}
\caption{Examples of the typical dynamical phases observed in confined termites. Top view of the arenas and showing, from left to right, disorder, clustering, milling, and an interesting coexistence of milling and clustering.}
\label{fig:figure1}
\end{center}
\end{figure}

Termites are, rightfully, biological active matter. 
They form groups of interacting individuals, and such interactions result in collective behaviours which translate into spatio-temporal emergent patterns \cite{miramontes1996, Miramontes2014,paiva2021}. 
Here we explore the various emergent spatial patterns in individuals confined in arenas. These collective arrangements of termite individuals are commonly found in the field \cite{grasse1937recherches, grasse1951sociotomie} as well as in the lab (as observed here)(Figure~\ref{fig:figure1}). Individuals in termite groups can be seen in the field and in the lab (i) wandering around, forming no observable pattern, or (ii) they cluster themselves in a dense and almost immobile pack, or (iii)  mill about in a circular movement.  Given the similarities between termite spatial organization and non-living active matter, we hipothesize that these collective arrangements of termites ({\it i.e.} disorder, clustering, milling) are parts of the same continuum, rather than behaviours originated from an independent selective pressure. In order to inspect this hypothesis, we parameterized phase changes in termite workers movement patterns in order to explore how interactions at a local smaller scale may control these changes, that is, the scale of the individual. Such a parameterization allowed us to inspect whether these changes lead to a change of one of these collective behaviours into another. 

Besides a disordered phase, where termites are basically moving uncorrelated, collective-directed motion and stable spatial clusters in groups of workers can be clearly identified. 
The disorder is observed at the beginning of an experiment when termites are just deposited in the arenas and they are exploring the new surroundings. 
In this sense, the disordered phase is like the transient period of a dynamical system when the basis of attraction is being explored before arriving into the phase space attractor.

Collective-directed motion is known as milling or collective vortex where self-propelled particles rotate spontaneously in circular motion around a common centre. 
In termites, the formation of milling was first reported, and named ``carrousel'',  by Grassé and others in {\it B. natalensis} and {\it B. belicosus} workers. 
Grassé noted that following a big perturbation, workers self-organize into a convoy, rotating in a circle for a very long time. 
He mentioned further that it was possible to often observe the spontaneous emergence of a second concentric circle of rotating termites but with the rotation direction inverted. (see \cite{grasse1937recherches,grasse1951sociotomie, grasse1986termitologia} and references therein). 

\begin{figure}[h!]
\begin{center}
\includegraphics*[width=\textwidth]{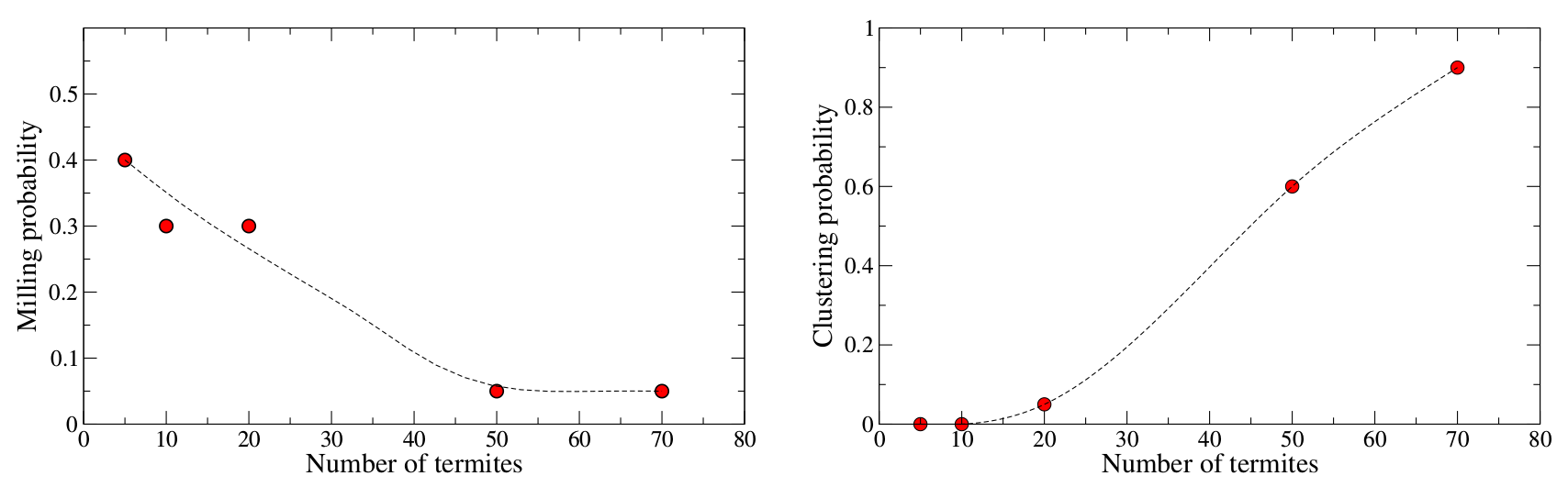}
\caption{\label{fig:figure2} Role of the termite density in the emergence of two collective behaviours. Left is the probability of observing milling when density is varied in the containers. As the number of workers increases, the probability of milling decays because of overcrowding. At right, the probability of observing clustering increases along with the density. Each point in the plots corresponds to an average of 20 independent samples, always in  $95$mm diameter arenas and about 1-hour experiments. Dashed lines are a guide for the eyes only.} 
\end{center}
\end{figure}

Circling behaviours highly similar to termite milling are also observed in a variety of organisms: ants \cite{schneirla1944unique}, caterpillar circles, bat doughnuts, amphibian vortex, duck swirl, and fish torus are just a few examples \cite{delcourt2016collective}. 
Milling is a kind of universal spatial phenomenon of abstract self-propelled particles, not necessarily alive, and is also present in simple computer models with or without spatial confinement, borders or walls \cite{levine2000self, d2006self, lukeman2009conceptual, costanzo2022effect, costanzo2018spontaneous, cambui2018milling, costanzo2019milling}. In nature, when a scouting termite finds a profitable source of food, it returns to its nest laying an odour trail to be subsequently followed by its foraging nestmates. As these foragers follow these marks and find the resource, they reinforce this trail so that to keep track of the resource location. At a first glance, this could explain the milling observed by us in the lab and by Grassé \cite{grasse1986termitologia} in the field. However, unprofitable sources of food do not stimulate scout termites to lay trails. Thence, providing that these field or lab circular paths do not lead to any food, it seems unwarranted to claim a role for trail pheromones in forming milling. Moreover, even if pheromones would play a role in maintaining the milling for some time, they would not explain the inception of milling. It seems thence reasonable to suspect that termite milling has some roots on the causes of milling spatial phenomena observed in abstract self-propelled particles which are not alive.

Clusters in termites are groups of individuals in close proximity, often engaged in body-to-body interactions through antennations but with the body barely moving. Clustering formation resembles a gas-liquid transition \cite{Cheng_2016} where initial free and uncorrelated moving termites condense (get trapped) into clusters by means of social interactions that act as the attractive and condensation force.  
Termites are polar-like particles, as they have distinct heads and tails, so disorder-order changes are expected. 
In fact, they interact through a combination of steric repulsion and alignment interactions.  It is reasonable to expect that the clustering phase is more likely to be observed at higher densities. 
In these conditions, most of the workers stay in the cluster, but a few may escape executing long walks in the arena. 
These roles (trapped and escaped) are interchanged in a way that most of the termites will get trapped again giving up the free walks at some point\cite{paiva2021}. It is remarkable that the step-size statistics of this process which includes mostly short movements but also few long walks is self-similar (Lévy-like) \cite{Miramontes2014,paiva2021}. 

In this contribution we explore several dynamical phases in the behaviour of confined termites. We identify disorder, clustering and milling. We aim to characterize these phases and their changes in order to achive a deep understanding of self-propelled particles in active matter both in living and artificial systems where close-range contacts and collisions are important.

\section{Methods}

Termite workers ({\it Cornitermes cumulans}) were collected in the grounds of the UFSJ (Universidade Federal de S\~ao Jo\~ao Del-Rei ) at its  ``Alto Paraopeba'' campus, in the municipality of Ouro Branco, Minas Gerais, Brazil. Termites were then taken to an acclimatised laboratory as described in detail in \cite{Miramontes2014}. Experimental arenas consisted of a glass Petri dish upside down over a filter paper. Within the arena, a variable number of termite workers was inserted and their movements were recorded continuously from above, with a video camera for 55 minutes (Sony FDR AX-53 4K). Termite trajectories were captured and digitised at a sample rate of one point every 0.333 s with an automatic video-tracking software (for more details see Supplementary Information of the reference \cite{paiva2021}.
No obstacles or food were present in the arenas. The resulting time series contained $x$, $y$ spatial coordinates used for numerical analysis.

\section{Results and Discussion}

Figure \ref{fig:figure1} displays images of the arena, illustrating experiments that represent the typical phases observed during the temporal evolution of the experiments. 
Each panel of Figure \ref{fig:figure1} pertains to a different experiment. 
The panels qualitatively illustrate the existence of three phases, with images of the disordered, clustering, and milling phases from left to right. 
We witnessed time series containing a single phase along all the observation time, as well as experiments exhibiting a change between phases, and occasionally the coexistence of multiple phases. 
The rightmost panel of Figure \ref{fig:figure1} presents an image exemplifying a scenario of phase coexistence (clustering and milling). 
One should emphasize that these phases can last from a few minutes to more than an hour.

\subsection{Role of density}

An individual may be recruited to a particular dynamical behaviour (Figure \ref{fig:figure2}) as a function of the number of individuals already committed to that behaviour \cite{deneubourg1989collective}, a phenomenon commonly known as social facilitation \cite{zajonc1965social}. 
However, this may be a non-linear response since the autocatalytic recruitment of individuals can be slowed down or halted when saturation happens beyond a given number or density of individuals. 
In ants, it was shown that a transition from a chaotic to a periodic state is a function of the density of individuals in a nest \cite{miramontes1995order}. 
In termites, it was shown that density can regulate individual survival \cite{miramontes1996nonlinear} or may enhance mating encounters by changing speed according to the density \cite{mizumoto2020termite} and in robots, it may induce the formation of aggregation clusters \cite{deblais2018boundaries}. 
Even vehicle traffic exhibits a phase transition as a function of density \cite{chowdhury2004self}. 
All this is because these are collective behaviours that depend on the number of participating individuals.

In our experiments (Figure \ref{fig:figure2}), we noticed that milling in these termites requires a low number of individuals to emerge. 
As a matter of fact, when the number is high and so is the density, we notice a low probability of milling formation. 
This seems counter-intuitive at first because high-density states would mean more social interactions to strengthen collective action; however high density disrupts the formation of coherent spatial structures simply because of the lack of space due to the finite boundaries of the containers. 
On the opposite, the emergence of clustering needs a large number of participating individuals since it is an aggregative process. (Figure \ref{fig:figure2}).

\subsection{Time series}

Termite walking and their movement patterns have been the subject of recent studies ranging from their anomalous diffusion properties \cite{Miramontes2014, paiva2021} to the dynamics of turning angles \cite{jeon2012simulation, mizumoto2019adaptive}. 
It is precisely the analysis of turning angles that allows us to explore additional aspects of the emergence of behavioural phases.
From the time series recordings, simple parameter measurements such as angular position were made and then used further to estimate a number of quantities as follows. 

\begin{figure}[h]
\begin{center}
\includegraphics*[width=0.8\textwidth]{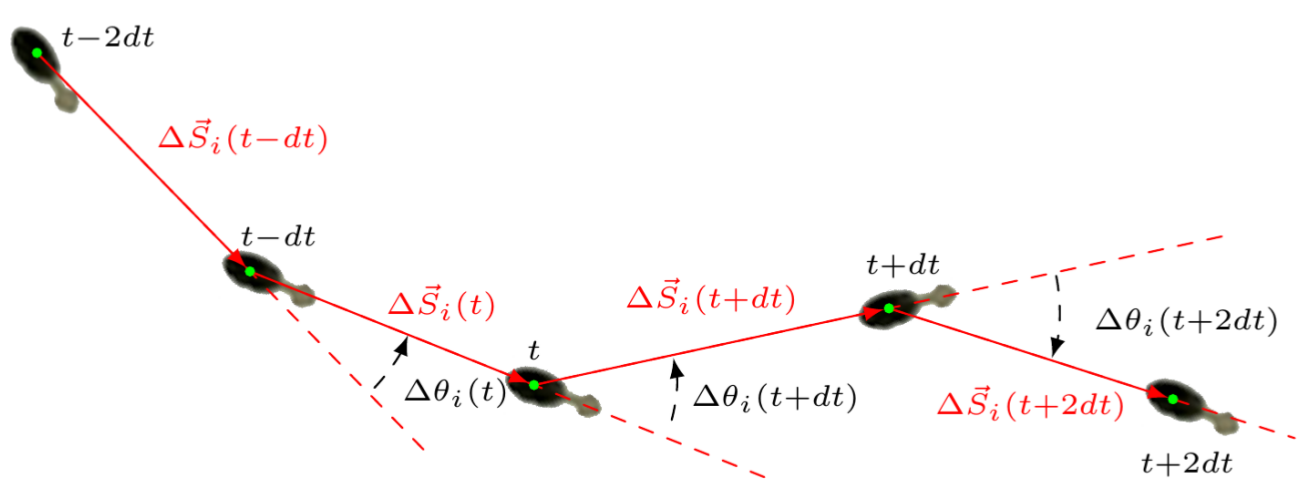}	
\caption{Schematic illustration of a sequence of steps of a termite between times $t-2dt$, $t-dt$, $\ldots$ $t+2dt$. 
        The red segments indicate the displacements between two consecutive positions. 
        The red dashed lines indicate the turning angle or angular displacement, between the direction of a step and the next one. 
        The dashed black arrows show the rotation direction. }
\label{fig:figure3}
\end{center}
\end{figure}

At each time step $t$, the $i$-termite position $\vec{S_i}(t)=(x_i(t),y_i(t))$ was recorded and the displacement was defined as

\begin{equation}
 \Delta \vec{S_i}(t)= \vec{S_i}(t) - \vec{S_i}(t-dt),
\end{equation}

\noindent as shown in Figure \ref{fig:figure3}. 
The velocity was obtained using $\vec v_i(t)=\Delta \vec{S_i}(t)/dt$ where $dt$ is the interval between the two consecutive positions. 
The turning angle $\Delta\theta_i(t)$ $\in (-\pi,\pi)$
was defined as the angle variation between two consecutive displacements, as shown in Figure \ref{fig:figure3} and using the following formula:

\begin{equation}
\theta_i(t) = \theta_i(t-dt) + \Delta\theta_i(t),
\end{equation}

\noindent we have defined the angular position $\theta_i(t)$, a temporal measure of the displacement's direction, of the termite $i$. 

\begin{figure}[!ht]
\begin{center}
\includegraphics*[width=\textwidth]{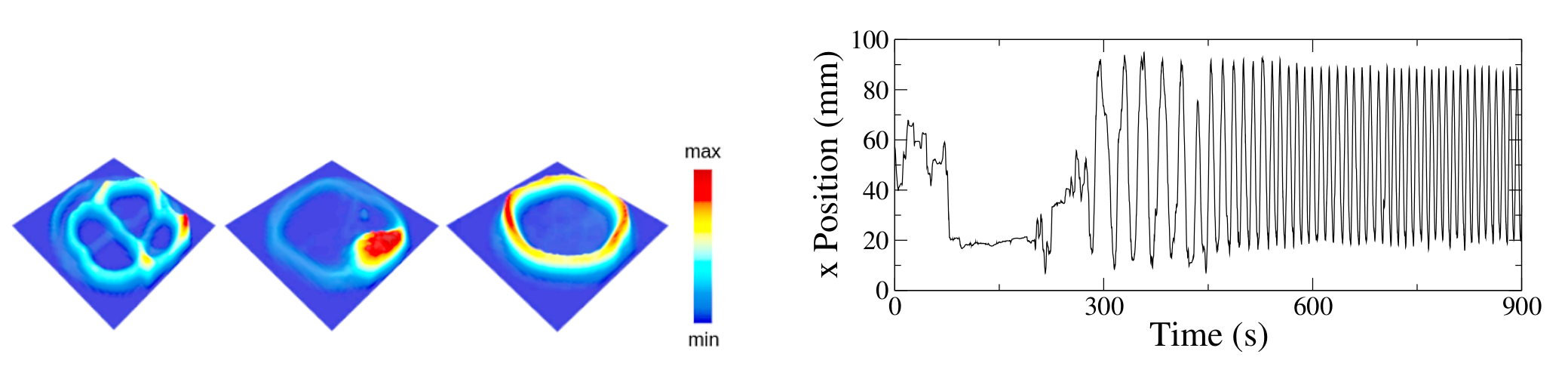}		
\caption{Typical temporal and spatial emergent patterns observed in confined termites. 
(Left) Accumulated activity in the arena as seen in a heat map; from left to right, disorder ($45$ termites, $105$mm-diameter arena)., clustering ($50$ termites, $95$mm-diameter arena)., and milling phases ($20$ termites, $100$mm-diameter arena). 
To build these plots, the arena area was divided into a square grid of size $40\times40$ units and the number of termites that entered and left each grid box during the experiment was recorded. 
(Right) Temporal evolution of the displacement $x$ spatial coordinate exhibiting three phases, for one termite in an $100$mm-diameter arena together with other $19$ termites.
At around $t=150$ seconds, there is an episode of clustering where the curve shows a plateau. 
At around 200 seconds there is a disordered phase when the stable plateau is broken and from $t\approx 450$, there is a milling phase exhibiting typical periodicity. }
\label{fig:figure4}
\end{center}
\end{figure}

Considering the observed phases, it is clear that the integrated trajectories of the termites exhibiting them are quite different. 
Therefore, it is important to examine quantitatively the spatial distribution of termites throughout the arena. 
To address this aspect, we defined a square that encloses the arena, splitting the area into a grid to determine the spatial distribution frequency. 
There are $40\times 40$ bins in a grid that divides this square. 
Throughout the experiment, we kept track of how often termites visited each grid cell. Note that only the cells inside the circular arena were visited.  
The resulting plots, for each behavioural phase, are presented as examples in Figure \ref{fig:figure4} left. 
In the disordered phase, several trajectories are visited. 
Meanwhile, in the clustering phase, the termites stay confined to a given region and, in the milling phase, a closed loop is more visited than other regions in the arena. 

\subsection{Displacement and turning angle fluctuations}

To quantitatively characterise the termite behaviour and the emerging phase, we consider the individual displacement time evolution obtained from the recorded videos. 
Initially, we aim to illustrate the behaviour for each phase by examining the temporal displacement in the $x$ spatial coordinate (displacement fluctuation). Each phase can be identified subjectively in a first approximation (see Figure \ref{fig:figure4}(right)). 
In the disordered phase, the displacement fluctuates quite significantly; in the clustering case, we observe very small or no fluctuations; and finally, in the milling phase, the displacement exhibits periodic behaviour. 
For convenience and without loss of generality, the $x$ position was analysed here using a fluctuation ratio $r$ given by: 

\begin{equation}
r=<x^2> / <x>^2.
\end{equation}

\begin{figure}[h]
\begin{center}
\includegraphics*[width=\textwidth]{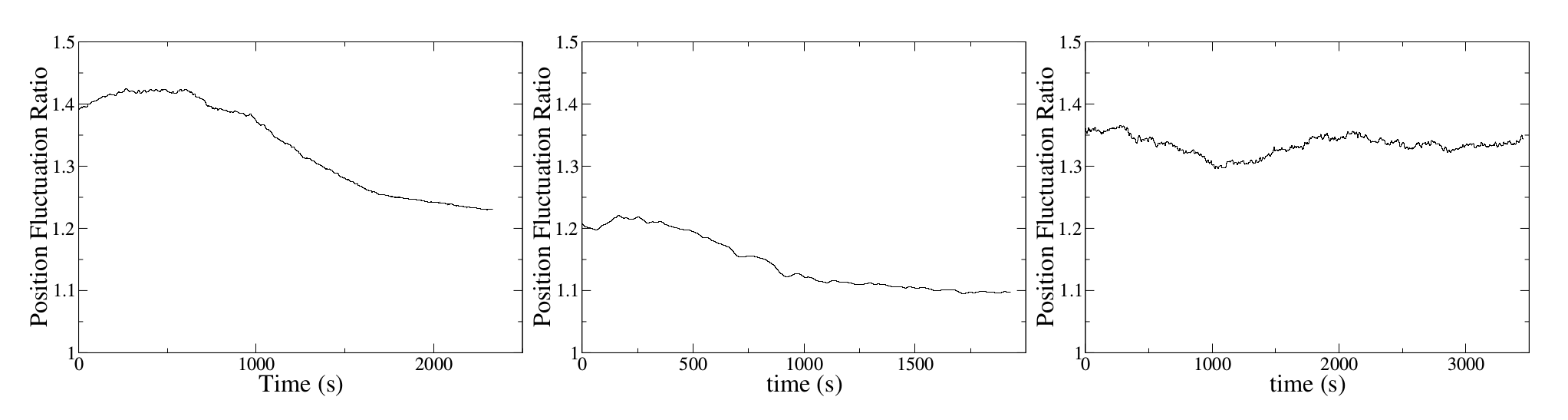}
\caption{\label{fig:figure5} Typical examples of a position fluctuation ratio analysis ($r$).  (Left and centre) Nine-time series containing 50006 points each were analysed and their $r$ was calculated using a moving average window of size $1.5\times10^4$. The average of all these points is shown as the black S-shaped curve that resembles a characteristic curve of a phase change. Left panel is a change from disorder to milling and the centre panel depicts a change from disorder to clustering. (Right) Seven-time series containing 48772 points each were analysed and their $r$  was calculated using a moving average window of size $1.5\times10^4$. The average is shown as the black curve exhibiting a nearly flat response. This is the case of disorder with no transition.}
\end{center} 
\end{figure}

Results show that time series where a change from disorder to milling is present can be spotted easily because they exhibit the typical S-shaped curve of a phase change (Figure \ref{fig:figure5}). 
On the other hand, series without milling exhibit a rather flat fluctuation ratio along time. 
When a change from disorder to milling or to clustering is present, the value of $r$ has the tendency to decay towards a lower value, suggesting a more coherent dynamical behaviour.

\begin{figure}[h!]
\begin{center}
\includegraphics*[width=\textwidth]{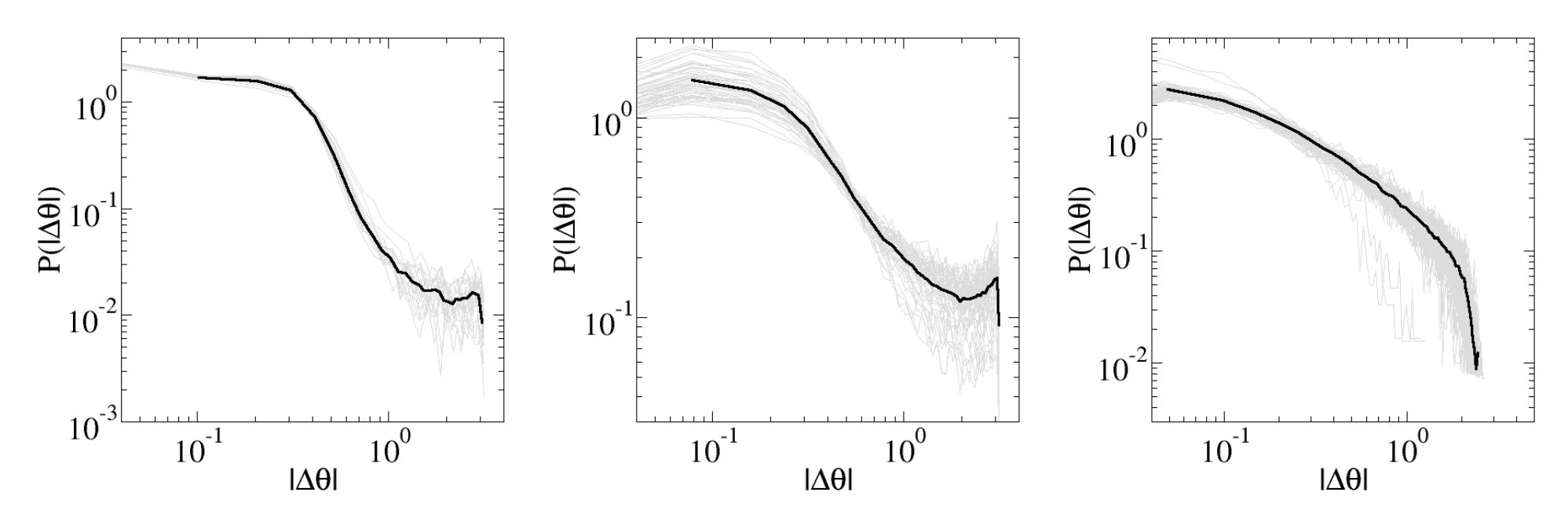}	
\caption{\label{fig:figure6} Typical plot examples of termite time series containing the probability distribution of angle fluctuations ($|\Delta\Theta|$) for different behavioural phases. 
Grey curves on each of them are individual termites in the arena and the black curve is the average of them. 
From left to right side: change from disorder to milling (20 time series containing 10866 points each), change from milling to clustering (50 time series containing 10374 points each) and disordered behaviour (78 time series with 2505 points each). 
Black S-shaped curve is obtained when there is a behavioural change. However in its absence, when there is just disordered behaviour, the average curve exhibits a simple decay response.}
\end{center}
\end{figure}

Another quantitative measure used to distinguish each phase and the changes from one phase to another is the turning angle. 
In particular, fluctuations of turning angles can be explored by means of a probability distribution of angle variations  ($|\Delta\Theta|$) as shown in the log-log plots in Figure \ref{fig:figure6} containing typical examples of time series with different behavioural phases on them. 
While this is a very simple analyses, it captures well the differences in the temporal behaviour when there are regime changes and when not.   

\subsection{Mean velocity and Hurst exponent}

The mean velocity of the termites increases when a change from clustering to disorder or from disorder to milling is observed. 
This could be expected, as the termites undergoing milling follow a path while in the disordered scenario, they are meandering through the arena. 
When they are in the clustering configuration, although a few termites break away from the cluster and walk around the arena, most of the termites are doing only small displacements around their position, so the mean velocity should be small.

\begin{figure}[!ht]
\begin{center}
\includegraphics*[width=\textwidth]{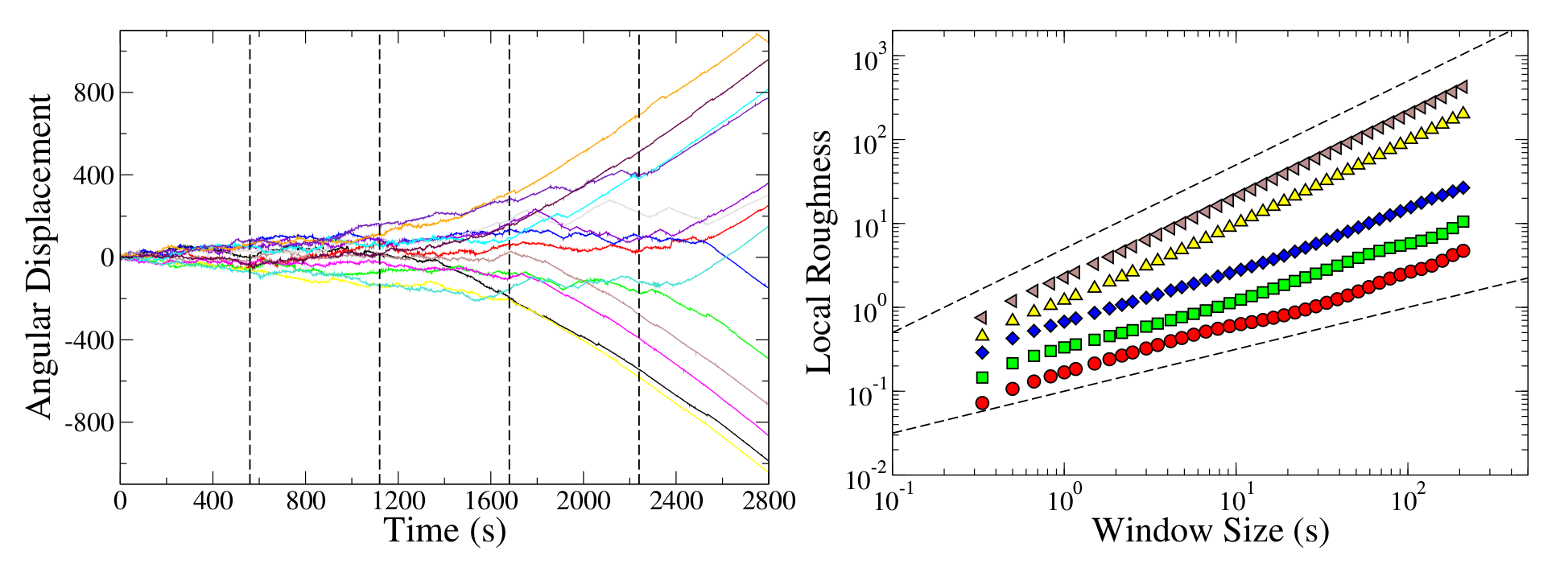}
\caption{\label{fig:figure7} Angular displacement (left panel) and local roughness (right panel) for an experiment in which the termites change their behaviour from disordered to milling after 30 minutes (there were $14$ termites in a $100$mm diameter arena). In left panel, each colour represents a different individual in the same arena. 
The vertical dashed lines in the left panel indicate five regions used to evaluate the local roughness (only five regions are shown for clarity). In right panel, each colour corresponds to the local roughness in one of the those regions (the first region in red circles, the second one in green squares, the third one in blue, the yellow triangles corresponds to the fourth region and the grey ones to the last region).
The dashed lines in the right panel, which represent power laws with exponents of 0.5 (bottom) and 1.0 (top), should serve as a visual guide.}
\end{center}
\end{figure}

The left panel of Figure \ref{fig:figure7} shows the time evolution of the angular position obtained for all termites in one experiment.
We analyse the curves obtained for the angular position considering the local roughness defined as the standard deviation of $\delta\theta_i(t)$ in relation to the mean inside a box of size $\varepsilon$: 

\begin{equation}
\displaystyle w^2(\varepsilon) = \left\langle \frac{1}{N_\varepsilon} \sum_t (\theta_i(t) - \langle\theta\rangle)^2\right\rangle_\varepsilon
\end{equation}

\noindent where $\langle\theta\rangle$ is the mean value of $\theta_i(t)$ and $N_\varepsilon$ the number of points inside the window of size $\varepsilon$. $\langle X \rangle_\varepsilon$ denotes a mean over the various windows of size $\varepsilon$. 
We consider ten sectors to measure the local roughness (the sectors are indicated by the vertical lines in the left panel of the Figure \ref{fig:figure7}). 

Through the scaling analysis of the local roughness ($w$) as the window size $\varepsilon$ increases, we observed that 

\begin{equation}
w(\epsilon) \sim \varepsilon^H,
\label{Hurst}
\end{equation}
\noindent where $H$ is the Hurst exponent, which provides information related to autocorrelation in time series \cite{HurstExp}.

$H$ values fall in the range [0,1], interpreted as follows.
A value $0.5 <H \leq 1$ indicates what is commonly termed `statistically persistent behaviour'; that is, whatever the past trend in the series, it is likely to
continue in the future, implying a strong degree of predictability and correlation. 
A value $0 <H \leq 0.5$ represents `anti-persistent behaviour' with low predictability.

\begin{figure}[ht]
\begin{center}
\includegraphics*[width=0.6\textwidth]{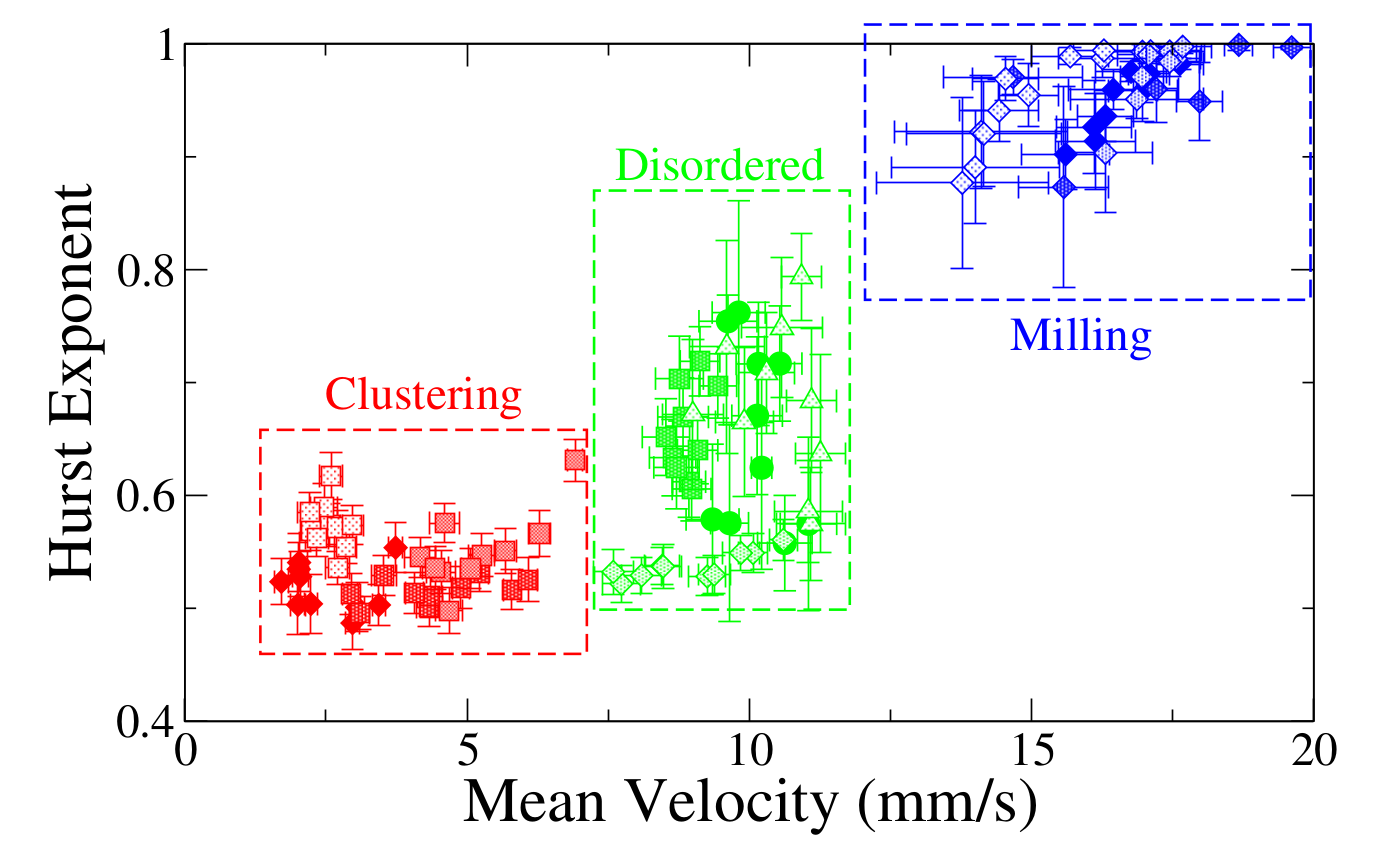}
\caption{\label{fig:figure8} A graph depicting the mean Hurst exponent against mean velocity. 
        Each point is an average from the measures of all termites in the arena in a time window. Same symbols correspond to the different time windows in the same experiment, and their colours were choose based on the emergent behaviour observed. One can see single phases well separated into defined regions as indicated by the dashed rectangles. Each experiment was divided into 10 time windows, within which we evaluated the Hurst exponent and the mean velocity. So, for a 55-minute video, this resulted in the analysis of ten short series, each lasting 5.5 minutes. This allows us to detect changes in emergent behaviour over time. Data are from 12 independent experiments that range from $3$ to $70$ termites, in arenas with diameters between $90$ and $105$mm.  }
\end{center}
\end{figure}

Since termites in milling are performing a persistent behaviour over a closed trajectory, it is expected that the Hurst exponent of their steps be close to $1$ while termites in the disordered state will be smaller. 
For each experiment, we split in ten equal parts all the time series (each one corresponding to one of the termites in the arena). 
Then, we averaged $H$ and the velocity $v$ over all termites in each time interval. 
In doing so, we obtain ten consecutive pairs of values ($<v>,<H>$) for each experiment. 
In Figure \ref{fig:figure8} these pairs are plotted and it becomes evident that the region occupied by them can be associated with each of the emergent behavioural phases.  
There is a clear pattern where milling is characterised by high values of $<H>$ and high-velocity values and clustering is characterised by low $<H>$ values and low-velocity values. 
Disordered behaviour is associated with intermediary values of both $<H>$ and $<v>$. 
It is interesting to note that, since our measurements were taken in a range of time windows, it allows us to identify the behavioural transitions across time.

\subsection{Momentum and standard deviation of turning angle}

\begin{figure}[!ht]
\begin{center}
\includegraphics*[width=0.8\textwidth]{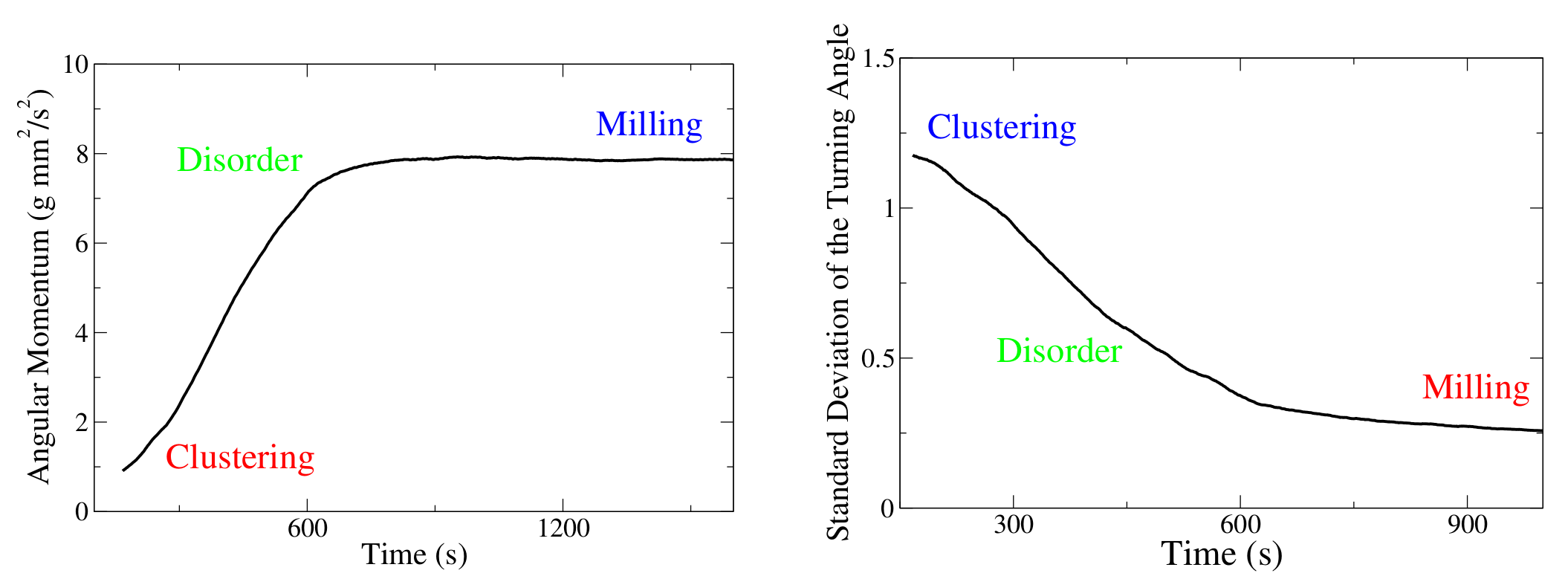}
\caption{Typical behaviours of the absolute value of the angular momenta (left) and of the standard deviation of the turning angle (right), averaged in all termites in the arena, in an experiment where the three phases were observed (in this particular experiment, there were $20$ termites in a $100$mm-diameter arena).}
\label{fig:figure9}
\end{center}
\end{figure}

To further characterise the collective behaviour of the termite groups and their behavioural phases we use two order parameters based on previous measures discussed elsewhere in simulation models and studies of schooling fish
\cite{Tunstrom2013,Couzin20021,Kolpas2007}. 
First, the rotation order parameter describes the rotation around the center of the arena for each time, it is defined as

\begin{equation}
O_L = \frac{ (1/N)\sum_{i=1}^N  \left| \vec{u}_i \times \vec{r}_i\right|} {\max\{\sum_{i=1}^N (1/N)\left| \vec{u}_i \times \vec{r}_i\right|\}},
    \label{eq:order_momentum}
\end{equation}

\noindent here, $\vec{u}_i$ and $\vec{r}_i$ are velocity and position vectors in relation to the center of the arena of the i-th termite, respectively. 
As in previous works, $O_L\in\{0,1\}$ by construction, where $0$ corresponds to no-rotation and $1$ to strong rotation about the center of the arena. 
The absolute values are important here because, in contrast to most of the organisms that perform milling, termites do not necessarily move all of them in the same direction: some of them can move in the clockwise direction while others are anticlockwise, changing directions eventually. 
As one can see in Figure \ref{fig:figure9}, when the termites are in the milling phase, one observes strong rotation. 
In the clustering phase there is a weak rotation and in the disordered phase, one observes intermediate values of the mean value of the absolute angular momenta. 
In the second-order parameter, we consider the standard deviation of the turning angle of the termites. 
In contrast to angular momenta, this parameter decreases as we go from clustering to disorder to milling, as can be seen in Figure \ref{fig:figure9}. 
The construction of this order parameter considers a normalisation by the maximum value in order to get $O_{\Delta\theta}\in\{0,1\}$.

\begin{figure*}[h]
\begin{center}
\includegraphics*[width=0.7\textwidth]{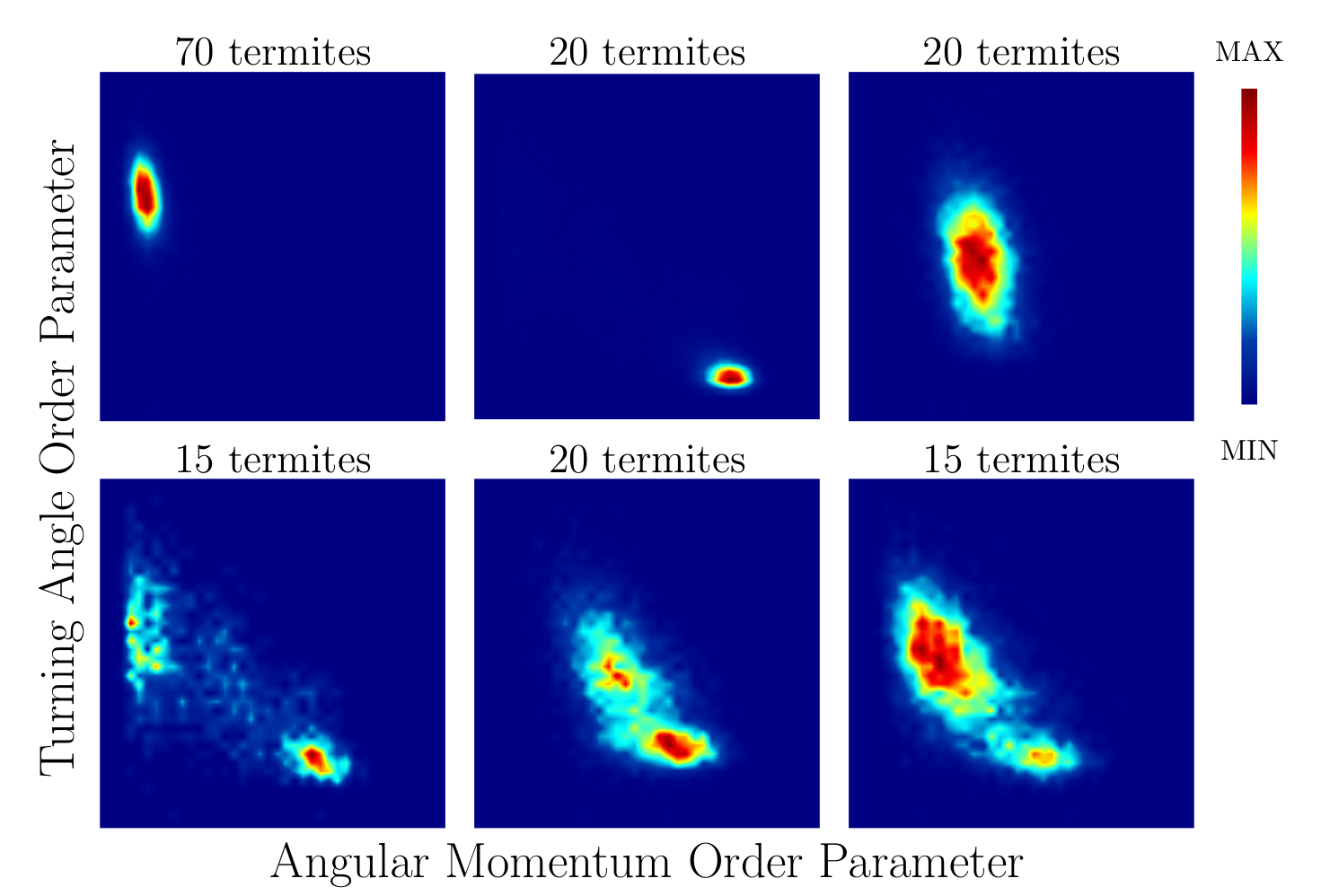}	
\caption{\label{fig:figure10}Density plots of the variance of the turning angle vs. angular momentum order parameters from six independent experiments, both variables in the interval $[0,1]$ by construction in each plot. 
The data show typical cases of each phase (top panels, from left to right: clustering, milling, and disordered) and also cases where behavioural transitions are observed (bottom panel, from left to right side: from milling to clustering, from disordered to milling, and from clustering to disordered to milling. Each plot was built using time series with between $10^4$ and $10^5$ data points . 
The experiments correspond to (right to left side, top to bottom): 70 termites in a 95-mm arena, 20 termites in a 100-mm arena, 20 termites in a 95-mm arena, 15 termites in a 100-mm arena, 20 termites in a 100-mm arena and 15 termites in a 90-mm arena. }
\end{center}
\end{figure*}

To demonstrate more clearly these dynamically stable states, we show in Figure  \ref{fig:figure10} a two-dimensional phase space spanned by the order parameter related to the variance of the turning angle $O_{\Delta \theta}$ and the one related to the angular momentum $O_L$. 
The mean turning angle of all termites analyzed is zero (data not shown), and values of $O_{\Delta \theta}$ close to $1$ are associated with a meandering behavior. 
Higher values of $O_L$ are associated with stronger rotation about the center of the arena. 
In this figure, we consider the proportion of time spent in different regions of this phase space, with red representing more time spent in a given region and blue the least time.

\noindent The top panels of the Figure \ref{fig:figure10}, from left to right, display typical cases where we observe clustering, milling, and disordered, during the entire trial. Some examples of the transition between different phases are shown in the bottom panels. 
Specifically, the changes from milling to clustering, disordered to clustering, and clustering to disordered to milling are displayed (from left to right, respectively).

Understanding the emergence of collective behaviour in active matter and the transitions from one state or phase to another from the interactions of many individual self-propelled constituents is challenging, especially in biological systems. 
In termites, under laboratory confinement, there are at least three detectable dynamical phases: disorder, clustering, and milling with transitions from disorder to clustering and disorder to milling. The explanations of what triggers these transitions remain elusive and largely unknown \cite{delcourt2016collective}. Density, velocity, alignment, and collisions (termites are blind \cite{eggleton2011introduction}) are factors worth exploring and we have produced in the present study a number of experiments aimed at identifying parameters that can characterise the phases and their changes). However, it is important to remember that milling (vortex behaviour) is ubiquitous to many organisms and artificial systems under very different spatial situations \cite{delcourt2016collective} and so the importance to provide new insights on termites is  relevant. The same applies to the other phases we describe here.  Grassé \cite{grasse1937recherches} could not point the causes of milling but he did point out that it was a natural event that occurred in the field. Our experimental setup was able to replicate, in the lab, these three field behaviours and that gave us confidence in drawing biological meaning from our assays. 

We hope that this study could be helpful for improving our understanding of diverse behavioural aspects of self-organised patterns and phase transitions in active matter including other social living species and artificial systems such as swarm models \cite{rauch1995pattern}, programmable robots \cite{thalamy2019survey} and engineering and interdisciplinary applications such traffic flows \cite{fourrate2006disordered}, smart aggregates \cite{saravanan2015comparative} and shape-memory materials \cite{lendlein2019bioperspectives}. 

\subsection*{Ethics} This study did not require ethical approval from any committee.

\subsection*{Dataccess} Time series containing examples of termite walking trajectories are 
available at the  FAIR-aligned Harvard Metaverse repository: \url{https://doi.org/10.7910/DVN/YJNXNQ}

\subsection*{Aucontribute} LRP and SGA: conceptualisation, laboratory experiments, formal analysis, software development. ODS and OM: formal analysis. All authors participated in writing the original draft and edited drafts.

\subsection*{Competing} No competing interests.

\subsection*{Funding} This study was financed by the Coordenação de Aperfeiçoamento de Pessoal de Nível Superior – Brasil (CAPES) – Finance Code 001, as well as the Minas Gerais State Foundation for the Support of Scientific Research (Fapemig), and the Brazilian National Council for Scientific Development (CNPq). OM was supported by UNAM PASPA – DGAPA. ODS holds a CNPq Fellowship \# 307328/2023-6 and SGA holds a CNPq Fellowship \# 311019/2021-8. 

\subsection*{Ack} OM thanks the Biological Physics Lab at UFSJR-Brazil and the Termitology Lab at UFV-Brazil 
for their hospitality during a research visit to them. Thanks to Silvio Ferreira, Ricardo Falcão, 
Gustavo Daudt, Dalson Oliveira and Daniel Ferreira for helping with the experiments. 
We also thank the free software community for the computational applications needed for data 
storage and manipulation, data analyses, image processing, typesetting, etc., through GNU-Linux/Debian, 
Ubuntu, Xubuntu, \LaTeX, BibTeX, Python, Grace, openCV, Overleaf, among others. 
This is contribution \# 88 from the Lab of Termitology at UFV (\url{http://www.isoptera.ufv.br}) 
and \#01 from the Biological Physics Lab at UFSJ.


\end{document}